\documentclass[sigconf, language=french,
language=german, language=spanish, language=english]{acmart}

\AtBeginDocument{%
  \providecommand\BibTeX{{%
    \normalfont B\kern-0.5em{\scshape i\kern-0.25em b}\kern-0.8em\TeX}}}

\copyrightyear{2022} 
\acmYear{2022} 
\setcopyright{rightsretained} 
\acmConference[CIKM '22]{Proceedings of the 31st ACM International Conference on Information and Knowledge Management}{October 17--21, 2022}{Atlanta, GA, USA}
\acmBooktitle{Proceedings of the 31st ACM Int'l Conference on Information and Knowledge Management (CIKM '22), Oct. 17--21, 2022, Atlanta, GA, USA}
\acmISBN{978-1-4503-9236-5/22/10}
\acmDOI{10.1145/3511808.3557142}

\usepackage{etoolbox}
\makeatletter
\patchcmd{\maketitle}{\@copyrightpermission}{
   \begin{minipage}{0.3\columnwidth}
     \href{https://creativecommons.org/licenses/by/4.0/}{\includegraphics[width=0.90\textwidth]{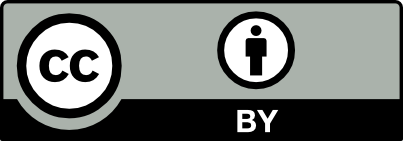}}
   \end{minipage}\hfill
   \begin{minipage}{0.7\columnwidth}
     \href{https://creativecommons.org/licenses/by/4.0/}{This work is licensed under a Creative Commons Attribution International 4.0 License.}
   \end{minipage}
  
   \vspace{5pt}
}{}{}

\makeatother

\usepackage{balance}
\usepackage{soul}
\usepackage{url}
\usepackage[utf8]{inputenc}
\usepackage{graphicx}
\usepackage{amsmath}
\usepackage{amsthm}
\usepackage{booktabs}
\usepackage{algorithm}
\usepackage{algorithmic}
\usepackage{multirow}
\usepackage{latexsym}
\usepackage{subfigure}
\usepackage{color}
\usepackage{placeins}

\begin{document}

%%
%% The "title" command has an optional parameter,
%% allowing the author to define a "short title" to be used in page headers.
\title{PEMP: Leveraging Physics Properties to Enhance Molecular Property Prediction}

\author{Yuancheng Sun}
\affiliation{%
  \institution{Institute of Automation, Chinese Academy of Sciences}
  \institution{University of Chinese Academy of Sciences}
  \institution{Beijing Academy of Artificial Intelligence}
%   \institution{Institute for AI Industry Research (AIR), Tsinghua University}
%     \city{Beijing}
%   \country{China}
}
\email{sunyuancheng2021@ia.ac.cn}
\authornote{Work done during Sun and Chen's internship at AIR, Tsinghua University.}
\authornote{Both authors contributed equally to this paper.}

\author{Yimeng Chen}
\affiliation{%
  \institution{Academy of Mathematics and Systems Science, Chinese Academy of Sciences}
  \institution{University of Chinese Academy of Sciences}
%   \city{Beijing}
%   \country{China}
}
\email{chenyimeng14@mails.ucas.ac.cn}
\authornotemark[1]
\authornotemark[2]

\author{Weizhi Ma}
\affiliation{%
  \institution{Institute for AI Industry Research (AIR), Tsinghua University}
%   \city{Beijing}
%   \country{China}
}
\email{mawz@tsinghua.edu.cn}

\author{Wenhao Huang}
\affiliation{%
  \institution{Beijing Academy of Artificial Intelligence}
%   \city{Beijing}
%   \country{China}
 }
\email{whhuang@baai.ac.cn}

\author{Kang Liu}
\affiliation{%
  \institution{Institute of Automation, Chinese Academy of Sciences}
  \institution{University of Chinese Academy of Sciences}
%   \institution{Beijing Academy of Artificial Intelligence}
%   \city{Beijing}
%   \country{China}
}
\email{kliu@nlpr.ia.ac.cn}

\author{Zhiming Ma}
\affiliation{%
  \institution{Academy of Mathematics and Systems Science, Chinese Academy of Sciences}
%   \institution{University of Chinese Academy of Sciences}
%   \city{Beijing}
%   \country{China}
}
\email{mazm@amt.ac.cn}

\author{Wei-Ying Ma}
\affiliation{%
  \institution{Institute for AI Industry Research (AIR), Tsinghua University}
%   \city{Beijing}
%   \country{China}
}
\email{maweiying@air.tsinghua.edu.cn}

\author{Yanyan Lan}
\affiliation{%
  \institution{Institute for AI Industry Research (AIR), Tsinghua University}
%   \city{Beijing}
%   \country{China}
}
\email{lanyanyan@tsinghua.edu.cn}
\authornote{Corresponding Author.}

%%
%% By default, the full list of authors will be used in the page
%% headers. Often, this list is too long, and will overlap
%% other information printed in the page headers. This command allows
%% the author to define a more concise list
%% of authors' names for this purpose.
\renewcommand{\shortauthors}{Yuancheng Sun et al.}

%%
%% The abstract is a short summary of the work to be presented in the
%% article.
\begin{abstract}
  A clear and well-documented \LaTeX\ document is presented as an
  article formatted for publication by ACM in a conference proceedings
  or journal publication. Based on the ``acmart'' document class, this
  article presents and explains many of the common variations, as well
  as many of the formatting elements an author may use in the
  preparation of the documentation of their work.
\end{abstract}

\begin{abstract}
    Molecular property prediction is essential for drug discovery. In recent years, deep learning methods have been introduced to this area and achieved state-of-the-art performances.
    However, most of existing methods ignore the intrinsic relations between molecular properties which can be utilized to improve the performances of corresponding prediction tasks. 
    % However, the applicability of these data-driven methods has been limited due to the data scarcity in this domain. In this paper, we propose a new approach named PEM to handle this problem. In comparison to existing works, we leverage the fact that molecular physics and chemistry properties are intrinsically related, which has been revealed by previous physics theory and physical chemistry studies. Specifically, we enhance the training of .......
    In this paper, we propose a new approach, namely Physics properties Enhanced Molecular Property prediction (PEMP), to utilize relations between molecular properties revealed by previous physics theory and physical chemistry studies. Specifically, we enhance the training of the chemical and physiological property predictors with related physics property prediction tasks. We design two different methods for PEMP, respectively based on multi-task learning and transfer learning. Both methods include a model-agnostic molecule representation module and a property prediction module. 
    % In particular, the representation module is model-agnostic, making our methods flexible to work with various molecular embedding models, such as graph neural networks and pretrained graph Transformer. 
    In our implementation, we adopt both the state-of-the-art molecule embedding models under the supervised learning paradigm and the pretraining paradigm as the molecule representation module of PEMP, respectively. Experimental results on public benchmark MoleculeNet show that the proposed methods have the ability to outperform corresponding state-of-the-art models.
\end{abstract}

%%
%% The code below is generated by the tool at http://dl.acm.org/ccs.cfm.
%% Please copy and paste the code instead of the example below.
%%


%%
%% Keywords. The author(s) should pick words that accurately describe
%% the work being presented. Separate the keywords with commas.

%% A "teaser" image appears between the author and affiliation
%% information and the body of the document, and typically spans the
%% page.
% \begin{teaserfigure}
%   \includegraphics[width=\textwidth]{sampleteaser}
%   \caption{Seattle Mariners at Spring Training, 2010.}
%   \Description{Enjoying the baseball game from the third-base
%   seats. Ichiro Suzuki preparing to bat.}
%   \label{fig:teaser}
% \end{teaserfigure}

%%
%% This command processes the author and affiliation and title
%% information and builds the first part of the formatted document.
\maketitle
% \pagestyle{plain}
% % \pagestyle{empty}

\section{Introduction}
% beginning from AI for drug design
Drug discovery is a time-consuming and expensive process, involving typical timelines of 10–20 years and costs that range from 0.5 billion to 2.6 billion US dollars~\cite{paul2010improve,avorn20152}. Over the past few years, interest has grown in applying artificial intelligence techniques to accelerate the drug discovery process~\cite{schneider2020rethinking}.

% importance of MPP, the prediction of bio/chemisty properties
Molecular properties prediction (MPP) is one of the most central tasks for computer-assisted drug discovery and has attracted much attention in the AI research society. Concerned properties for molecules include their physics, chemistry, biology, and physiology properties~\cite{wu2018moleculenet}. Among these properties, the
chemistry and physiology properties are especially focused, like solubility, lipophilicity, and membrane permeability, because they indicate the drug-likeliness and affect the pharmacokinetics process~\cite{hamidi2013pharmacokinetic, zhong2017admet}. Accurate predictions on these properties are critical for early-stage drug selection and also vital for the efficiency of molecular generation, where the target function is usually defined by an MPP model.
% especially those related with the Pharmacokinetics process~~\cite{}.

% Deep learning for molecular descriptions, its success
% Problems of this strategy, only self-supervised
Previous studies have shown the effectiveness of deep learning methods for the prediction of chemistry and physiology properties~\cite{gilmer2017neural,xiong2019pushing,hu2019strategies,rong2020self}. 
Typically, a graph neural network is utilized to encode each molecule, and a property predictor is then used to output the prediction score. In the learning process, both parameters of encoder and predictor are estimated by the supervised learning method based on labeled training data. In recent years, people have designed different graph neural networks, such as MPNN~\cite{gilmer2017neural} and AttentiveFP~\cite{xiong2019pushing}, and shown superior prediction results as compared with traditional predictors based on hand-crafted molecule representations.

% depend on the stereochemistry of molecules
However, labeled data for chemical and physiological properties are usually very scarce, due to the expensive experimental process, which limits the ability of the powerful supervised deep learning methods. Recently, the pretraining paradigm has been used in this problem, and gain some improvement by leveraging large-scale unlabeled data to obtain a better molecule representation~\cite{hu2019strategies,rong2020self}. However, the data scarcity issue remains in their downstream training~\cite{wang2021property}.

To tackle this problem, we resort to investigating useful factors related to chemistry and physiology properties. Fortunately, we found important domain knowledge that some calculable physics properties can affect or inform physics-chemical and physiology properties that are concerned in drug discovery. For example, physical chemistry theory~\cite{peter2018atkins} has shown that dipole moment and isotropic polarizability, which are two molecule polarity properties, directly affect inter-molecular forces and thus inform the solubility and lipophilicity of the molecules.
% Further, the inter-molecular forces is a primary factor that determines solubility and lipophilicty.,
% They are theoretically related with the solubility and lipophilicity of the molecules which are determined by inter-molecular forces~\cite{}. 
Such relation has also been verified by some empirical research in physical chemistry properties~\cite{rubino1987cosolvency, abraham1999correlation}. Moreover, the polarity of molecules has been found to influence the permeability of molecules for biological membranes in the principle of drug design~\cite{zhong2017admet}. These domain knowledge inspires us to leverage physical properties to enhance the prediction of chemistry and physiology properties. One benefit of this approach is that physical property data are usually easier to obtain and also more reliable than chemical data, because physical properties are intrinsically determined by molecules themselves, and could be directly calculated by some mature computation methods~\cite{cramer2013essentials}, instead of wet-lab experiments.\par

Based on the above domain knowledge, we propose a \textbf{P}hysics properties \textbf{E}nhanced \textbf{M}olecular \textbf{P}roperty prediction (PEMP) approach, to enhance the training of the chemical and physiological property predictors with related calculable physics property prediction tasks. Specifically, we design two methods for PEMP, respectively based on multi-task learning and transfer learning.
% using physics properties to enhance chemical and physiological properties prediction tasks 
The proposed two methods are capable to work with various representation modules used in existing MPP models, such as graph neural networks and pretraining models. In the training process, the prediction losses of auxiliary tasks and the target task are simultaneously optimized in the multi-task training method; while in the transfer learning method, the model is first pretrained on auxiliary tasks in a multi-task way, then transferred and fine-tuned on target tasks.

In our experiments, we take AttentiveFP~\cite{xiong2019pushing} and GROVER~\cite{rong2020self}, which are representative molecule representation models in graph neural networks and pretraining models respectively, as example representation modules to demonstrate our model architecture. We adopt the prediction of physics properties in the QM9 dataset~\cite{ramakrishnan2014quantum} as the auxiliary tasks to enhance the prediction performance of chemistry and physiology properties, such as solubility, lipophilicity, permeability, and toxicity.
Experimental results on six MPP datasets in MoleculeNet show that the proposed methods consistently outperform corresponding state-of-the-art baselines. Furthermore, we conduct ablation studies to show the effectiveness of our selected auxiliary tasks.

Our main contributions are summarized as follows:
\begin{itemize}
    \item the investigation of existing physical chemistry theory and empirical physical chemistry studies about the implicit relations between different molecular properties;
    \item  the proposal of PEMP to utilize physics properties to enhance the chemistry and physiology properties and the design of two methods based on multi-task learning and transfer learning;
    \item extensive experimental studies on six MPP datasets in MoleculeNet, which show the superiority of our proposed methods with different backbone representation modules, as against state-of-the-art models.
\end{itemize}

 \begin{figure*}[t]
  \centering
  \subfigure[Basic]{
      \includegraphics[scale=0.48]{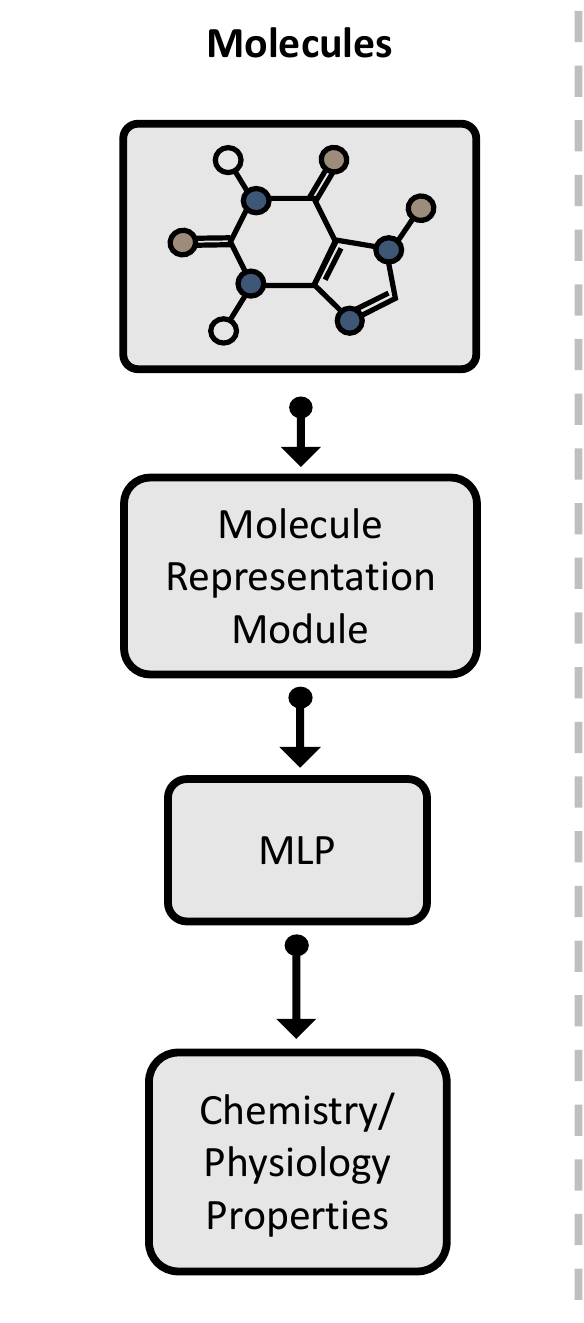}
      \label{subfig:basic}
  }
  \subfigure[PEMP-MTL]{
      \includegraphics[scale=0.48]{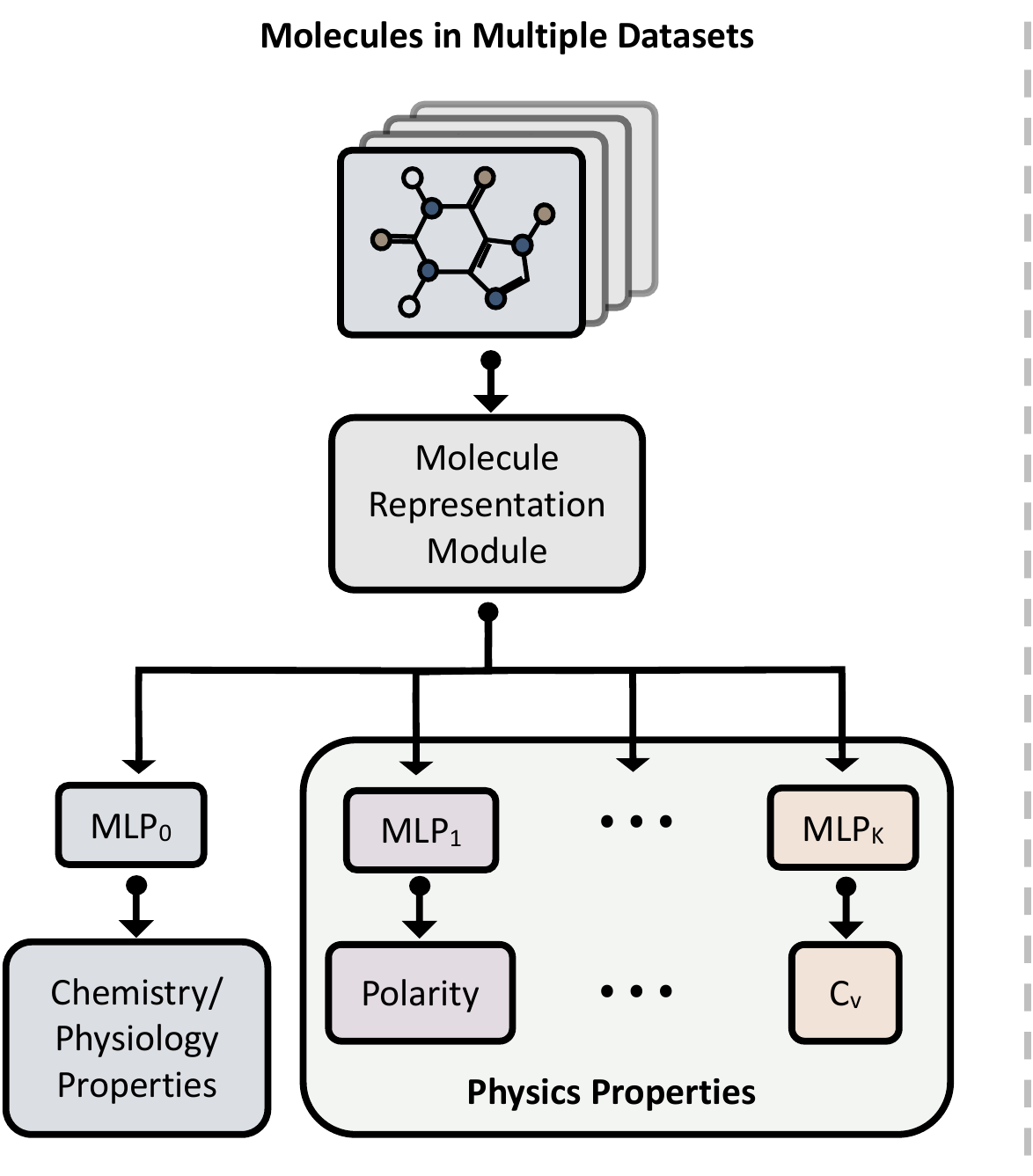}
      \label{subfig:mtl}
  }
    \subfigure[PEMP-TransL]{
      \includegraphics[scale=0.48]{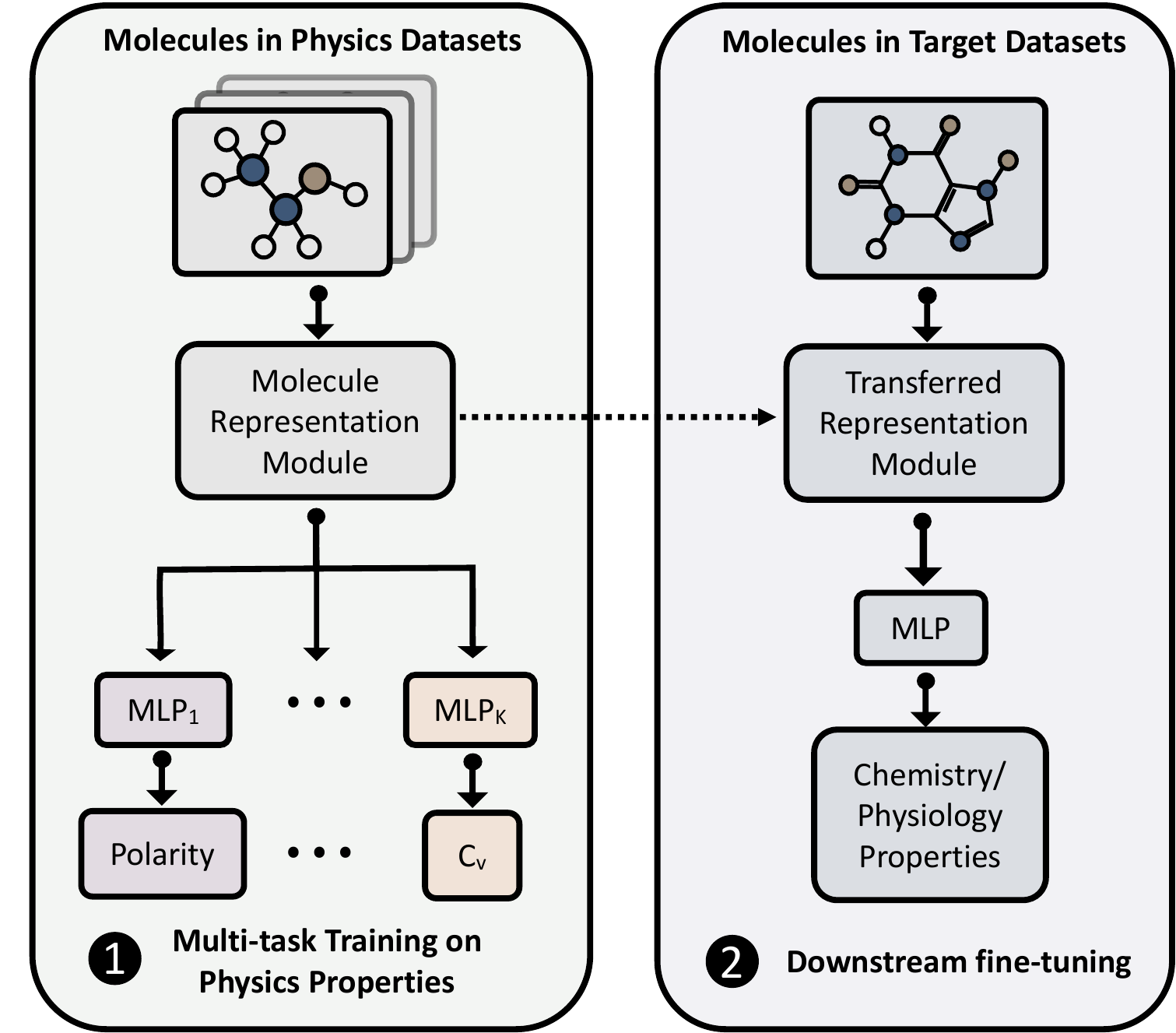}
      \label{subfig:tl}
  }
    \caption{(a) Basic approach for comparison. Predictive models for different target properties are trained separately. (b) Multi-task Learning based PEMP. The model are jointly trained on datasets of target and auxiliary properties. Loss is calculated on labeled properties of each sample during training. (c) Transfer Learning based PEMP. The representation module is initialized via the multitask training on auxiliary physics properties, then updated with the task-specific predictor during fine-tuning on target datasets.\label{fig:method}}
\vspace{-10pt}
\end{figure*}

\section{Related Works}
In this section, we introduce some related works on molecular property prediction, including graph neural network based methods, pretraining methods, and some recent works in bio-informatics using multi-task learning.
\label{sec:related}
%\paragraph{MPP with Graph Neural Networks.}

Encoding molecules into learnable vectors lies in the center of cheminformatics. Traditional methods use hand-crafted descriptors to encode the neighbors of atoms in the molecule into a fix-length vector, namely molecule fingerprints~\cite{rogers2010extended}. In recent years, Graph Neural Network (GNN) has drawn increasing attention and shown promising results in many fields, including MPP. Steven \textit{et al.}~\shortcite{kearnes2016molecular} conduct convolutions on simple encodings of the molecular graph---atoms, bonds, distances, etc.--- and make greater use of molecule graphs to learn molecule representations. With attention mechanism thriving, people propose Graph Attention Networks to learn nonlocal intramolecular interactions, which can draw chemical insights directly from data beyond human perceptions~\cite{xiong2019pushing,zhang2021fragat}. Glimmer \textit{et al.}~\shortcite{gilmer2017neural} first show that message passing neural networks with appropriate message, update, and output functions are effective for predicting molecular properties. Johannes \textit{et al.}~\shortcite{klicpera2020directional} associate the messages between atoms with a direction in coordinate space, alleviating the limitation that previous frameworks do not consider the directional information from one atom to another.
    
Since molecule properties are intrinsic and should stay unchanged to translations and rotations in 3D space, i.e. SE(3)-(in)equivariance, there has been a trend in building equivariant GNNs to learn more robust and knowledgeable molecule represetations~\cite{han2022geometrically}. By projecting features into orthogonal spaces and designing special update layers, Fuchs \textit{et al.}~\shortcite{fuchs2020se} and Satorras \textit{et al.}~\shortcite{satorras2021n} successfully introduced group theory into molecule modelling and achieved SOTA at that time. Schütt \textit{et al.}~\shortcite{schutt2021equivariant} propose polarizable atom interaction networks to reduce model size and inference time, but the architecture only works for predicting tensorial properties and molecular spectra.\par

Besides the straightforward supervised-learning paradigm, researchers also explore to pretrain a model on large-scale datasets and then transfer it to the target property with a small size of training data. Following this paradigm, Ye \textit{et al.}~\shortcite{ye2018integrated} collect a number of labeled datasets about toxicity which contains 500,000 molecules with 157 targets, pretrain a model on it, and transfer to the target set with only 1104 FDA approved drugs. Further, people turn to unlabelled data and utilize self-supervision to do pretraining. For example, Hu \textit{et al.}~\shortcite{hu2019strategies} proposes a policy specifically for graph by masking nodes/edges and then predicting graph contexts. More recently, people are working on adding domain knowledge to those structure-based models to improve the representations. Rong \textit{et al.}~\shortcite{rong2020self} tries to involve chemical knowledge by proposing a semantic motif task to predict the functional groups in masked molecules. Song \textit{et al.}~\shortcite{song2020communicative} and Yang \textit{et al.}~\shortcite{yang2021deep} integrate 3D conformations to obtain the molecule representations.

Above approaches rely on labeled property data to conduct the training process. However, such data are usually very scarce~\cite{wu2018moleculenet,brown2019guacamol,townshend2020atom3d}. In the tasks of molecule-protein interaction prediction, Multi-task learning (MTL) has been adopted widely to tackle this problem. For example, an MTL method on 19 cellular and biochemical assays which are relevant to the target proteins is adopted in~\cite{dahl2014multi}. Liu \textit{et al.}~\shortcite{liu2021multi} proposes a multi-task learning method to utilize task relations encoded by a protein-protein interaction graph. Wang \textit{et al.}~\shortcite{wang2021property} learn a task relation graph to encode the difference between targets but under the few-shot learning setting.
Besides, Wenzel \textit{et al.}~\shortcite{wenzel2019predictive} conduct straightforward MTL on the prediction of each ADME property (e.g. metabolic clearance) across different species.
Our work is different from these related works since we consider the chemistry and physiology property prediction tasks on molecules. To the best of our knowledge, the idea of leveraging physics properties to enhance the predictions of chemistry and physiology properties is novel in MPP.

\begin{table*}
% \vskip 0.15in
\begin{center}
\begin{small}
\caption{Statistics of datasets. \label{tab:datasets}}
\begin{tabular}{l|c|c|c|c|c|c|c}
\toprule
Datasets & QM9 ($\mu,\alpha,gap,C_v,U0$) & Freesolv & Esol & Lipo & BBBP & Clintox & Tox21 \\
\midrule
\#molecules & 133885 & 642 & 1128 & 4200 & 2039 & 1478 & 7831 \\
\#targets & 5 & 1 & 1 & 1 & 1 & 2 & 12 \\
\midrule
Task Type & Regression & \multicolumn{3}{c|}{Regression} & \multicolumn{3}{c}{Classification} \\
Category & Quantum Physics & \multicolumn{3}{c|}{Physical Chemistry} & \multicolumn{3}{c}{Physiology} \\
\midrule
Metric & MAE & \multicolumn{3}{c|}{RMSE} & \multicolumn{3}{c}{AUC-ROC} \\
\bottomrule
\end{tabular}
\end{small}
\end{center}
% \vskip -0.1in
\end{table*}

\section{Motivation}
\label{sec:motivation}
% Nice :)
Typical molecular property prediction tasks include predicting the physics, chemistry, and physiology properties of molecules. Among them, the most important MPP tasks are chemistry and physiology ones. The reason is that these molecule properties can indicate the drug-likeliness or influence the pharmacokinetics process, which is critical in the whole drug discovery pipeline. 

As the chemistry and physiology properties are usually measured in wet-lab experiments, only a limited number of labeled data are collected, especially for physiology properties that need in vivo studies~\cite{wu2018moleculenet}. As a result, supervised learning or finetuning on these properties faces the problem of data scarcity. To tackle this problem, we turn to consider the physics properties.

Quantum physics properties of molecules, such as dipole moment, isotropic polarizability, are direct parameters that characterize the electronic structure of molecules and can be accurately computed with multiple theoretical methods~\cite{cramer2013essentials}. For example, the magnitude of the dipole moment of a molecule is the product of the length of the vector pointing from its negative charge center to the positive one and the quantity of charge. In physical chemistry theory, the electronic properties of molecules determine the molecular interactions~\cite{peter2018atkins}. For example, intermolecular forces including dipole-dipole interactions, dipole–induced dipole interactions, and dispersion interactions all depend on the dipole moments of molecules. As molecular interactions govern the behavior of molecules in multiple chemical and physiological processes, we believe that physics properties are informative for predicting these target properties.\par

In fact, these conjectures have been proven in recent physical chemistry theory and empirical studies, which guide us to select proper physics properties to enhance the prediction of chemistry and physiology properties. Dipole moment and isotropic polarizability are two properties on the polarity of molecules. They have been shown to correlate with the solubility and lipophilicity of the molecules by empirical research~\cite{abraham1999correlation}. The polarity of molecules also affects the permeability of molecules for biological membranes~\cite{zhong2017admet}. 
The HOMO-LUMO gap affects the reactivity of molecules in biochemical process~\cite{kumer2019simulating}, which potentially indicates the toxicity of molecules~\cite{mekenyan1994qsars}. 
The internal energy and heat capacity of molecules indicate the strength of the intermolecular force and the flexibility of the scaffold of molecules~\cite{peter2018atkins}, which affects the solubility, lipophilicity, metabolism, and toxicity of molecules~\cite{zhong2017admet}. Therefore, we select five physics properties, i.e.~dipole moment, isotropic polarizability, HOMO-LUMO gap, internal energy, and heat capacity as the auxiliary properties, to facilitate the prediction of the solubility, lipophilicity, permeability, and toxicity of molecules, which corresponds to the following datasets in MoleculeNet ~\cite{wu2018moleculenet}: Freesolv, Esol, Lipophilicity, BBBP, Clintox and Tox21. Detailed information of these datasets is given in Section\ref{sec:setup}

\begin{table*}
% \vskip 0.15in
\begin{center}
% \begin{small}
\caption{Experimental results on various datasets. Methods with the notation $_\text{TransL}$/$_\text{MTL}$ means the method is enhanced with our transfer learning framework or multi-task learning framework. The best performances in typical deep learning and pretraining paradigm are underlined and the SOTA values are bolder. The numbers in brackets are the standard deviation for split seeds. `-' means the method is unable to handle this task. `$\dagger$' indicates the model's results are listed as \protect\cite{rong2020self}. \label{tab:main}}
\begin{tabular}{l|ccc|ccc}
\toprule
{\multirow{2}{*}{\textbf{Methods}}} & \multicolumn{3}{c|}{\textbf{Regression (Lower is better)}} & \multicolumn{3}{c}{\textbf{Classification (Higher is better)}} \\
& \textbf{Freesolv} & \textbf{Esol} & \textbf{Lipo} & \textbf{BBBP} & \textbf{ClinTox} & \textbf{Tox21} \\
\midrule
$\text{SchNet}^{\dagger}$ & $3.215_{(0.755)}$ & $1.045_{(0.064)}$ & $0.909_{(0.098)}$ & $0.847_{(0.024)}$ & $0.717_{(0.042)}$ & $0.767_{(0.025)}$ \\
$\text{MPNN}^{\dagger}$ & $2.185_{(0.952)}$  & $1.167_{(0.430)}$ & $0.672_{(0.051)}$ & $\underline{0.913}_{(0.041)}$ &
$0.879_{(0.054)}$ & $0.808_{(0.024)}$ \\
AttentiveFP & $2.092_{(0.695)}$ & {${0.833}_{(0.120)}$} & $0.660_{(0.039)}$ & $0.845_{(0.044)}$ &
$0.909_{(0.066)}$ & $0.828_{(0.029)}$ \\
\midrule
\textbf{$\text{AttentiveFP}_\text{MTL}$} & $\underline{1.811}_{(0.630)}$ & $\mathbf{\underline{0.790}_{(0.123)}}$ & $0.636_{(0.017)}$ & ${0.912}_{(0.025)}$ & $\mathbf{\underline{0.936}_{(0.031)}}$ & $0.826_{(0.011)}$ \\
\textbf{$\text{AttentiveFP}_\text{TransL}$} & {${1.918}_{(0.526)}$} & $0.835_{(0.086)}$ & {$\underline{0.602}_{(0.035)}$} & 
{${0.894}_{(0.032)}$} & $0.908_{(0.081)}$ & $\mathbf{\underline{0.852}_{(0.039)}}$\\
\midrule
\midrule
$\text{N-GRAM}^{\dagger}$ & $2.512_{(0.1390)}$ & $1.100_{(0.160)}$ & $0.876_{(0.033)}$ & $0.912_{(0.013)}$ & 
$0.855_{(0.037)}$ & $0.769_{(0.027)}$ \\
$\text{Hu}\; \textit{et al.}^{\dagger}$ & - & - & - & $0.915_{(0.040)}$ & $0.762_{(0.058)}$ & $0.811_{(0.015)}$   \\
% GROVER LARGE & $1.544_{(0.397)}$  & $0.831_{(0.120)}$ & $0.560_{(0.035)}$ & $0.940_{(0.019)}$ & 
%$0.925_{(0.013)}$ & $0.819_{(0.020)}$ \\
GROVER & $1.898_{(0.367)}$  & $1.066_{(0.191)}$ & $0.573_{(0.031)}$ & $0.893_{(0.032)}$ &  $0.901_{(0.018)}$ & $\underline{0.826}_{(0.013)}$ \\
% tanghan & $1.868_{(0.386)}$  & $0.969_{(0.118)}$ & & $0.919_{(0.030)}$ &  $0.780_{(0.070)}$ & \\
\midrule
% \textbf{OursMTL} & $2.030_{(0.418)}$ & $0.829_{(0.337)}$ & $0.715_{(0.169)}$ & 
% $0.920_{(0.003)}$ & $0.883_{(0.039)}$ & $0.773_{(0.042)}$  \\
\textbf{$\text{GROVER}_\text{MTL}$} & $2.134_{(0.385)}$ & $\underline{0.821}_{(0.100)}$ & $0.625_{(0.032)}$ & 
$0.903_{(0.025)}$ & $0.884_{(0.045)}$ & $0.740_{(0.078)}$  \\
% \textbf{OursTransL} & $1.641_{(0.356)}$ & $0.860_{(0.123)}$ & $0.541_{(0.027)}$ & 
% $0.933_{(0.022)}$ & $0.912_{(0.032)}$ & $0.820_{(0.030)}$  \\
\textbf{$\text{GROVER}_\text{TransL}$} & 
$\mathbf{\underline{{1.659}}_{(0.413)}}$  & $0.864_{(0.152)}$ & 
$\mathbf{\underline{0.545}_{(0.036)}}$ & $\mathbf{\underline{0.928}_{(0.034)}}$ & $\underline{0.909}_{(0.049)}$ & $0.818_{(0.036)}$  \\
\bottomrule
\end{tabular}
% \end{small}
\end{center}
% \vskip -0.1in
\end{table*}

\begin{table*}
% \vskip 0.15in
\begin{center}
% \begin{small}
\caption{Ablation studies with respect to different auxiliary physics properties. \label{tab:ablation}}
\begin{tabular}{l|cc|cc}
\toprule
{\multirow{2}{*}{\textbf{Methods}}} & \multicolumn{2}{c}{\textbf{TransL}}& \multicolumn{2}{c}{\textbf{MTL}} \\
& \textbf{Esol} & \textbf{BBBP} & \textbf{Esol} & \textbf{BBBP} \\
\midrule
GROVER (no physics)  & $1.066_{(0.191)}$ & $0.893_{(0.032)}$ & $1.066_{(0.191)}$ & $0.893_{(0.032)}$   \\
\midrule
$\text{PEMP}_{\text{w/o} \, \mu \, \text{etc.}}$ & $0.920_{(0.103)}$ & $0.914_{(0.044)}$ 
& $0.877_{(0.109)}$  & $0.905_{(0.024)}$  \\
$\text{PEMP}_{\text{w/o gap etc.}}$ & $0.911_{(0.132)}$ & $0.915_{(0.028)}$  
& $0.899_{(0.116)}$   & $\mathbf{0.922}_{(0.018)}$  \\

$\text{PEMP}_\text{5 properties}$ & $\mathbf{0.864}_{(0.152)}$ & $\mathbf{0.928}_{(0.034)}$ & $\mathbf{0.821}_{(0.100)}$ & $0.903_{(0.025)}$ \\

\bottomrule
\end{tabular}
% \end{small}
\end{center}
% \vskip -0.1in
\end{table*}

\section{PEMP}
\label{sec:pem}
% This section presents our proposed PEM architecture 
In this section, we propose a novel approach called \textbf{P}hysics properties \textbf{E}nhanced \textbf{M}olecular chemistry \textbf{P}roperty prediction (PEMP). PEMP consists of two separate methods, namely the Multi-task Learning (MTL) method and the Transfer Learning (TransL) method, and both of them are able to introduce the related physics property prediction tasks for the target property prediction. Specifically, the MTL method aims to improve the generalizability of neural molecule representations by jointly learning physics and biochemistry properties of molecules, and the TransL method aims to provide a more contextualized molecular embedding and a more efficient training method. More information of the two methods will be discussed in section Section\ref{sec:pem}.

The goal of PEMP is to introduce knowledge from related tasks to the prediction of target molecular property. In particular, as analyzed in Section 3, we use MTL and transfer learning to leverage knowledge on physical properties of molecules into the chemistry property prediction in this study. 

% is the main body of the model and learns
The overall model architecture of the two frameworks is described in Figure~\ref{fig:method}.
The structure of PEMP consists of two main components: 1) the molecule representation module, which generates neural representations for the input molecule; 2) the property prediction module, which contains task-specific fully-connected layers to make independent predictions on multiple properties. We will first discuss about typical molecule representation modules in Section \ref{subsec:Encoder}, and elaborate on the two methods of property prediction module in Section \ref{subsec:MTmethod} and Section \ref{subsec:TLmethod}.

\subsection{Molecule Representation Module}
\label{subsec:Encoder}
The molecule representation module consists of a parameterized neural network, which encodes input molecules into neural representations as the inputs of the prediction module. Denote the input molecule in dataset $\mathcal{D}$ as $M^\mathcal{D}=\{m_1,m_2,...,m_N\}$, where $m_i$ is the input form of molecule $i$. Denote $\text{Enc}(\cdot, \theta)$ as the representation module parameterized by $\theta$, we have $\mathbf{e}_i = \text{Enc}(m_i, \theta)$, 
% \begin{equation}
%     \mathbf{e}_i = \text{Enc}(m_i, \theta)  \nonumber
% \end{equation}
where $\mathbf{e}_i \in \mathbb{R}^{d}$ is a $d$-dimensional representation of $m_i$.

Since the molecule representation module is designed to be model-agnostic, PEMP can be applied with any molecule representation learning model.
In our work, we adopt two different existing molecule representation models to implement $\text{Enc}(\cdot,\theta)$, namely AttentiveFP and GROVER, as the molecule representation module for PEMP. AttentiveFP is a supervised graph neural network using the attention mechanism, it not only models atoms and their neighbors but also uses the graph attention mechanism to characterize non-local effects at intra-molecular level. Compared to AttentiveFP, GROVER tries to apply the idea of pretraining on molecules by building a graph-based Transformer with its self-attention building blocks replaced by specially designed GNN.

Above two models have shown remarkable performance in modeling molecules, and we would like to demonstrate that PEMP can help such architectures make further progress. 

Following the molecule embedding module, we can build prediction modules upon it and use ground-truth annotation of molecular properties to supervise the training of the whole model. The following two sections talk about the two methods proposed for PEMP prediction module.

\subsection{Multi-Task Learning based PEMP}
\label{subsec:MTmethod}
In order to leverage task relations for prediction, the most intuitive method is to integrate multiple classifiers of different tasks through a jointly MTL framework (shown in Figure~\ref{subfig:mtl}). 
In Multi-task based PEMP, we use Multi-Layer Perceptron (MLP) to build separate classifiers for physics and chemistry properties, where different MLPs share a common molecule representation from the representation module. Suppose $\mathcal{D}^0$ is the dataset of a target property. $\{\mathcal{D}^1, \mathcal{D}^2, \dots, \mathcal{D}^K\}$ are datasets on $K$ different physics properties. 
For any $k \in \{0, 1, \dots, K\}$, denote $\mathcal{D}^k = \{(m^k_i, p^k_i)\}_i$, where $m^k_i$ is the input form of molecules, and $p^k$ is the labeled property. Denote $\text{MLP}_{k}(\cdot, \phi_k)$ as the prediction module for the $k$-th property parameterized by $\phi_k$, the prediction $\hat{p}_i^k$ is calculated as:
\begin{equation}
        \hat{p}_i^k = \text{MLP}_k(\mathbf{e}_i^k, {\phi}_k)  \nonumber
\end{equation}
In PEMP-MTL, the loss function are similar to existing multi-task learning methods, while the main difference is that our choice of auxiliary tasks are based on physics and physical chemistry studies. The loss function of PEMP-MTL is defined as a weighted sum of losses on the $K+1$ tasks:
\begin{align}
        \mathcal{L}_{MTL} = \sum_{i_0} {\ell_{0}(\hat{p}^0_{i_0}, p^0_{i_0})} +
        \sum_{k=1}^K {w_k} \sum_{i_k} {\ell_k (\hat{p}^k_{i_k}, p^k_{i_k})} \nonumber
\end{align}
where $l_0, l_1, \dots, l_K$ are task specific loss functions measuring the discrepancy between ground-truth label $p$ and the estimation $\hat{p}$. $\{w_1, w_2, \dots, w_K\}$ are hyper-parameters adjusting the weights of $K$ auxiliary physics properties.\par

Note that the used molecules do not necessarily have labels for all the tasks. During training, the model takes inputs from different datasets, and the loss is only calculated on labeled properties for each instance. As for the problem that datasets on different properties vary in sample sizes, we cyclically input the small dataset into the model before the model is fully trained on the large dataset for an epoch.

\begin{table*}
\caption{Statistics of hyper-parameters when finetuning the GROVER backbone. \label{tab:finetune}}
% \vskip 0.15in
\begin{center}
\begin{small}
\begin{tabular}{l|l|l}
\toprule
Hyper-parameter & Description & Range \\
\midrule
batch\_size & the input batch size & 32 \\
max\_lr & the max learning rate of Noam learning rate scheduler. & [0.0001,0.0002,0.001]\\
init\_lr & the initial learning rate ratio of Noam learning rate scheduler. & 10 \\
& The real initial learning rate is max\_lr/init\_lr & \\

final\_lr & the final learning rate ratio of Noam learning rate scheduler. & [2,5,10] \\
& The real final learning rate is max\_lr/final\_lr &\\
dropout & dropout ration & [0.2,0.1,0.05] \\
dist\_coff& GROVER encode molecules on node- and edge-view, &[0.15,0.10,0.05]\\
&this parameter controls how the two encodings are joined & \\
ffn\_num\_layer & the number of MLP layers & [2,3]\\
ffn\_hidden\_size & the hidden size of MLP layers & [700,1300]\\
\bottomrule
\end{tabular}

\end{small}
\end{center}
% \vskip -0.1in
\end{table*}

\subsection{Transfer Learning based PEMP}
\label{subsec:TLmethod}
% \paragraph{Trans}
% In PEMP-MTL, the learned representations are usually endpoint-specific, which means when we want to predict another property, the model has to be retrained from scratch. Further, considering the disparity of sample sizes, MTL costs much more on the quantum physical tasks than the target chemical task, which may cause a waste of resources. Thus, we propose another framework PEMP-TransL based on transfer learning, which gives a solution to this non-efficiency by providing contextualized molecular embeddings for downstream tasks.
In PEMP-MTL, the learned representations are usually endpoint-specific, which means when we want to predict another property, the model has to be retrained from scratch. Further, due to the disparity of sample sizes between chemical datasets and quantum physics dataset, MTL costs much more resources on processing the loss on auxiliary physics tasks than target chemistry tasks. Thus, we propose another framework PEMP-TransL based on transfer learning, which gives a solution to this non-efficiency by providing contextualized molecular embeddings for downstream tasks.

As is illustrated in Figure~\ref{subfig:tl}, the implementation of a transfer learning framework can be divided into the following two steps:
1) Pretrain a model through supervised tasks on physics properties to get general representations of molecular data; 2) Fine-tune the field-specific model on a chemistry or physiology task.

Compared to PEMP-MTL, PEMP-TransL introduces the auxiliary prediction tasks on physics properties into the molecule representation module, thus improving the flexibility of PEMP when dealing with different properties in the prediction module. 
In step 1, we conduct a multi-task pretraining on the auxiliary physics properties, where the model follows the same architecture of PEMP-MTL except that the prediction part for chemistry/physiology is excluded. Following the notations in Section \ref{subsec:MTmethod}, the pretraining process can be formulated as:
\begin{equation}
        \mathcal{L}_{TransL_1}  =
        \sum_{k=1}^{K} {w_k}{\sum_{i_k} {\ell_k(\hat{p}_{i_k}, p_{i_k})}}, 
        \nonumber
\end{equation}
and the optimal model we get can be defined as:
\begin{equation}
\label{equation:pretrain}
        \text{Enc}_{QM}^* =
        \operatorname*{arg\,min}_{\theta, \phi_{1}, \dots, \phi_{K}} \mathcal{L}_{TransL_1}   \nonumber
\end{equation}
Through pretraining, the model is able to well reflect the distribution of molecular properties and therefore gains knowledge that can be transferred to downstream tasks of chemistry properties. 
In step 2, the pretrained representation module is initialized with $\text{Enc}_{QM}^*$. The multi-task prediction module is replaced with a task-specific one, with the objective function:
\begin{equation*}
        \mathcal{L}_{TransL_2} =
        \sum_{i} {\ell_0} (\hat{p}^0_i, p^0_i)
\end{equation*}
During training, both the encoder and the predictor are updated under supervision of target labels, as is shown below:
% In fact, we can control the extent to which the physical knowledge are transferred by adjusting the learning rates of the two modules or simply freeze the representation module. In our work, the two modules are updated at the same learning rate.

\begin{equation}
\label{equation:trans}
        \text{Enc}_{Target}^*, \text{MLP}^* =
        \operatorname*{arg\,min}_{\theta, \phi} \mathcal{L}_{TransL_2}
        \nonumber
\end{equation}
% where the $\text{Enc}_{Target}^*$ is initialized by the optimal $\text{Enc}_{QM}^*$ obtained in step 1.

\section{Experimental Results}
In this section, we demonstrate our experiments on public benchmark datasets, to compare our methods with state-of-the-art molecular property prediction models.
\subsection{Experimental Setup}
\label{sec:setup}
We first introduce our experimental setup, including datasets and preprocessing, baselines, evaluation metrics, and empirical settings.

\subsubsection{\textbf{Datasets and Preprocessing.}} Experiments are conducted on two widely used public datasets/benchmarks: \textit{QM9} and \textit{MoleculeNet}. We briefly introduce the two datasets as follows:
\begin{itemize}
    \item \textbf{QM9}: QM9 dataset collects 130K molecules with up to 9 heavy atoms (C,O,N,F). It provides 13 physics quantum properties for each molecule. Over the past years, QM9 has been recognized as a ground truth for quantum chemistry research due to its high accuracy.
    \item \textbf{MoleculeNet}: Abbreviated as MolNet, it is a benchmark specially designed for testing machine learning methods of molecular properties. MolNet curates a number of dataset collections and categorizes them into four categories namely quantum mechanics, physical chemistry, biophysics, and physiology.
\end{itemize}

As is introduced in Section \ref{sec:motivation}, from MolNet we select Freesolv, Esol, Lipophilicity from physical chemistry properties and BBBP, Clintox, Tox21 from physiology properties as our target properties. Freesolv~\cite{mobley2014freesolv} is a small dataset which contains the hydration free energy of molecules. Esol~\cite{delaney2004esol} and Lipophilicity~\cite{gaulton2012chembl} are both properties indicating the water solubility or membrane permeability of molecules. BBBP~\cite{martins2012bayesian} documents whether a compound is able to penetrate the blood-brain barrier. Clintox~\cite{gayvert2016data} records drugs approved by the FDA and drugs that fails clinical trials for toxicity reasons. Tox21~\cite{"Tox21challenge"} is a public database measuring toxicity of compounds. As for auxiliary properties, we select the following quantum physical properties from QM9: dipole moment ($\mu$), isotropic polarizability ($\alpha$), HOMO-LUMO gap ($\epsilon_{gap}$), internal energy at 0 K ($U_0$) and the heat capacity at 298.15K ($C_v$). Table \ref{tab:datasets} gives statistics about these datasets.

We adopt the scaffold splitting~\cite{bemis1996properties} on the dataset other than the widely-used random split because the former method is more compatible and realistic in the context of bioinformatics and molecular property prediction. The ratio for train/validation/test is 8:1:1. For each dataset, we train the models for three replicas on three scaffold splitting with 0,1,2 as random seeds and report the means and standard deviations of the results. Specifically, when adopting GROVER as the backbone representation module, we conduct 3 runs for each seed and use the means as the results to alleviate the uncontrollable randomness in the PyTorch implementation of message passing network~\cite{rong2020self}.

\subsubsection{\textbf{Baselines and Evaluation Metrics.}} To evaluate the effectiveness of our proposed method, we deploy PEMP on two representative models, Attentive FP and GROVER, one stands for typical deep learning scheme and the other stands for the pretraining scheme. 
Besides, we separately compare the results of PEMP with baselines of the two paradigms: typical deep learning methods including SchNet~\cite{schutt2017schnet} which performs convolutions on graphs, MPNN~\cite{gilmer2017neural} which considers edge features during message passing, the naive AttentiveFP, and pretrained methods including N-Gram ~\cite{liu2018n} which embed the molecules by extending the N-gram approach in NLP from linear graphs (sentences) to general graphs (molecules), Hu et.al~\cite{hu2019strategies} which learn local and global molecule representations simultaneously, the naive GROVER. 
%We conducted experiments with official implementations of Attentive FP; for other baselines, as identical evaluation schemes were adopted, we referred to the reported performances in corresponding citations and left unreported entries blank.

For regression tasks, the root-mean-square-errors (RMSE) is adopted as the evaluation metric, and a lower score means better performance. For classification tasks, the averaged ROC-AUC is reported as the measure, and better performance will lead to a higher score. Both metrics follow the suggestions of the MoleculeNet benchmark~\cite{wu2018moleculenet}.

\subsubsection{\textbf{Empirical Settings and Hyper-parameters.}}
\label{subsec:empsetting}
In multi-task learning method, in order to handle the dataset size imbalance pointed out in \ref{subsec:Encoder}, we cyclically feed the small dataset of target properties into the model until the model is fully trained on the auxiliary dataset for an epoch. The model is trained for 5 epochs on the physics property dataset whatever the target property is. We conduct a grid search on weights $w_i$ among $\{1, 0.5, 0.1, 0.01\}$ and report the best results. The validation loss on target properties is used to select the best model. In transfer learning method, the model is first pretrained on the auxiliary dataset for 300 epochs and then fine-tuned on target properties for 100 epochs, and the best model is selected by validation loss. 

When adopting GROVER as the molecule encoder, which is a Transformer-style model, we adopt Adam optimizer and the Noam learning rate scheduler. For hyper-parameters, we perform grid-search on the validation set as suggested in ~\cite{rong2020self} and report the best results. Table \ref{tab:finetune} demonstrates the hyper-parameters we tuned. Except for \textit{batch\_size} and \textit{init\_lr}, the optional values of the hyper-parameters are separated with ',' in the square brackets. When adopting AttentiveFP as the encoder, we applied the hyper-parameter combination as the original implementation in its code base. \par

\subsection{Overall Performance}
Experimental results are summarized in Table~\ref{tab:main}. The best performances respectively in typical deep learning and pretraining paradigms are underlined, and the SOTA values on each dataset are in bold. 

Firstly, regardless of the backbone models we select, the two frameworks of PEMP can help both backbone models obtain better or comparable performances, which validates our motivation that it is vital to consider physical properties when predicting chemistry properties. PEMP boosts the performance with 6.9\% relative improvements on average over the original Attentive FP, and 9.1\% improvements over the original GROVER.
Secondly, although both methods are shown to be effective, PEMP-TransL shows more stable performance. Among the 12 experiments of TransL method, 9 achieve better results and the remaining 3 achieve comparable results.
Thirdly, the pretraining based methods outperform classical supervised methods on physics chemistry properties and membrane permeability but are defeated by typical deep learning models on toxicity. The contrast suggests the generic structural information embedded in pretrained models may not function better than the task-relation information when dealing with toxicity datasets.

\subsection{Impact of Physics Properties}
Since PEMP is proposed to utilize the physics properties of molecules to enhance the prediction of chemistry and physiology properties, people may be concern about how the choice of physics properties will affect the performances. In this section, we conduct some ablation studies to investigate this problem. According to the analysis in Section \ref{sec:motivation}, the polarity-related properties dipole moment ($\mu$) and isotropic polarizability($\alpha$) can affect the solubility of molecules, while HOMO-LIMO gap, internal energy ($U_0$) and heat capacity ($C_v$) are related to the scaffolds of molecules and can affect physiology properties. Therefore, we group the auxiliary physics properties into two groups and build two variants of PEMP, both using GROVER as backbone representation modules: (1) $\text{PEMP}_{\text{w/o} \, \mu \, \text{etc.}}$, which excludes $\mu$ and $\alpha$ yet keeps gap, $U_0$, $C_v$. (2) $\text{PEMP}_{\text{w/o gap etc.}}$, the variant model which is a counterpart of model(1). We select Esol to test solubility and BBBP to test physiology properties.

Table \ref{tab:ablation} shows the performance of different variants. The results of $\text{PEMP}_{\text{w/o}\, \text{physics}}$ (original GROVER) is listed as the baseline. From the results, We can obtain the following observations. With the assistance of all the 5 related quantum physics properties, PEMP performs better than both GROVER and the two variant models in most cases, which validates the usefulness of introducing related properties. Though excluding some of the auxiliary physics properties, the two variants still have the ability to outperform the plain baseline GROVER, indicating that both groups of selected properties are helpful.

\begin{table}
\label{tab:hyperablation}
\caption{Hyperparameter comparison of $\text{GROVER}_\text{TransL}$ \label{tab:hypercompare}}
\centering
% \begin{small}
\begin{tabular}{ll|c|c|c}
\toprule
\multicolumn{2}{c|}{\textbf{Hyper-parameters}} & Freesolv & Esol & BBBP \\
\midrule
{\multirow{3}{*}{dropout}} & 0.05 & 1.952 & 0.890 & 0.915  \\
& 0.1 & 1.909 & 0.881 & 0.916\\
& 0.2 & \textbf{1.868} & \textbf{0.877} & \textbf{0.919}\\
\midrule
{\multirow{2}{*}{ffn\_num\_layers}} & 2 & \textbf{1.862}& \textbf{0.874} & \textbf{0.917}  \\
& 3 & 1.957 & 0.891 & 0.915 \\
\midrule
{\multirow{2}{*}{ffn\_hidden\_size}} & 700 & \textbf{1.911} & 0.885 & \textbf{0.917 } \\
& 1300 & 1.908 & \textbf{0.880} & 0.916\\
\bottomrule
\end{tabular}
% \end{small}

\label{tab:hyperparam}
\end{table}

% \subsection{Hyper-parameter Analyses}

\subsection{Impact of Hyper-parameters}
A statistics of hyper-parameter grid search is displayed in Table \ref{tab:hypercompare}. For each row with a hyper-parameter fixed, the reported score is averaged over all the other hyper-parameters. For example, when the dropout value is set to 0.2, the reported value 1.868 for Freesolv is reported as an average of four experiments where \textit{ffn\_num\_layers} is set to 2 and 3, and \textit{ffn\_hidden\_size} is set to 700 and 1300, respectively. 

From the table, we can draw the conclusion that: The model with a high dropout rate and fewer parameters usually performs better. This is probably because when finetuning on GROVER, we are feeding thousands of data samples into a large model with millions of parameters, and the optimal model we obtain after 100 epochs' training has been overfitted on the train set. 
% Furthermore, we adopt the scaffold split strategy, which may lead to different distributions of molecules in the training and testing set. 
Therefore, a model with a higher dropout rate and fewer parameters can have better generalizability and lead to better performance on the test set. Considering that a large number of datasets in biology and chemistry have smaller sample sizes than those in machine learning~\cite{wu2018moleculenet,brown2019guacamol,townshend2020atom3d}, this result may be helpful on applying pretrained models or desiging large models for molecular tasks in the future. 

\section{Conclusions and Future Works}
In this paper, we propose a novel molecular property prediction approach called PEMP. %\textbf{P}hysics properties \textbf{E}nhanced \textbf{M}olecular chemistry property prediction approach.
The motivation comes from the fact that concerned chemistry and physiology properties are usually related to some physics properties, as proved by classical physical chemistry theory and some recent empirical research on physical chemistry properties. In PEMP, we enhance the prediction of molecular chemistry and physiology properties by auxiliary quantum physics properties. Specifically, two PEMP methods are designed respectively based on multi-task learning and transfer learning. PEMP is general to work with various molecular embedding methods, like existing deep learning paradigms and pretraining paradigms. In our experiments, we evaluate two versions of PEMP based on different embedding methods. The experimental results show that PEMP consistently outperforms state-of-the-art models.

This work complements the mainstream MPP models that make efforts on modeling molecule graphs and empirically shows that a proper learning framework can make more significant progress than advanced and complex neural network structures. These results verify the importance of domain knowledge in the study of AI4Science. In the future, we will try to refine the naive multi-task and transfer learning method in this work and we will explore PEMP on other related tasks like molecule generation and molecule interaction. Also, we are interested in integrating more domain knowledge of molecules to existing data-driven methods for better performances and understandings.

\begin{acks}
This work was supported by the National Key R\&D Program of China (2020AAA0105200), Vanke Special Fund for Public Health and Health Discipline Development, Tsinghua University (No.2022-1080053), Guoqiang Research Institute, Tsinghua University (2021-GQG1012), and the Key Research Program of the Chinese Academy of Sciences (Grant NO.ZDBS-SSW-JSC006). This research work was also supported by Youth Innovation Promotion Association CAS. We also want to thank the reviewers for their helpful suggestions.
\end{acks}

%%
%% The acknowledgments section is defined using the "acks" environment
%% (and NOT an unnumbered section). This ensures the proper
%% identification of the section in the article metadata, and the
%% consistent spelling of the heading.
% \begin{acks}
% To Robert, for the bagels and explaining CMYK and color spaces.
% \end{acks}

\bibliographystyle{ACM-Reference-Format}
\balance
\bibliography{cikm}

% \clearpage

\end{document}